\documentclass[10pt]{article}
\usepackage{subcaption, fontenc, contour}
\usepackage[export]{adjustbox}[2011/08/13]
\usepackage{graphicx,amssymb,amsfonts,amsmath, geometry, lipsum}
\usepackage{nameref,hyperref}
\usepackage{cite}
\usepackage{lineno}
\usepackage{anyfontsize}
\begin{document}
\pagenumbering{arabic}
 \textbf{{\fontsize{16pt}{1pt} \selectfont Acetylcholine-modulated plasticity in reward-driven navigation: a computational study. \newline }}
\newline
\\
Sara Zannone\textsuperscript{1},
Zuzanna Brzosko\textsuperscript{2},
Ole Paulsen\textsuperscript{2},
Claudia Clopath\textsuperscript{1}*
\\

\textbf{1} Department of Bioengineering, Imperial
College London, South Kensington Campus, London, United Kingdom, 
\textbf{2} Department of Physiology, Development and Neuroscience, Physiological
Laboratory, Cambridge, United Kingdom
\bigskip

* c.clopath@imperial.ac.uk \newline

\textbf {{\fontsize{14pt}{2pt}\selectfont Abstract  \newline}}

Neuromodulation plays a fundamental role in the acquisition of new behaviours. Our experimental findings show that, whereas acetylcholine biases hippocampal synaptic plasticity towards depression, the subsequent application of dopamine can retroactively convert depression into potentiation. We previously demonstrated that  incorporating this sequentially neuromodulated Spike-Timing-Dependent Plasticity (STDP) rule in a network model of navigation yields effective learning of changing reward locations. Here, we further characterize the effects of cholinergic depression on behaviour. We find that acetylcholine, by allowing learning from negative outcomes, influences exploration in a non-trivial manner that highly depends on the specifics of the model, the environment and the task. Interestingly, sequentially neuromodulated STDP also yields flexible learning, surpassing the performance of other reward-modulated plasticity rules. \\

\textbf {{\fontsize{14pt}{2pt}\selectfont Author summary  \newline}}

Animals have to learn to interact with their environment in order to survive. Learning is believed to be achieved by the brain through changes in the connection strength between neurons. This process is called synaptic plasticity. Neuromodulators are substances released in the brain that  modulate plasticity and correlate with behavioural changes. Experimental evidence suggests that feedback from the environment can be carried through neuromodulation. The neuromodulator dopamine is believed to act as a reward signal, while acetylcholine correlates with exploration and spatial learning. In our previous work, we investigated experimentally how dopamine and acetylcholine alter synaptic plasticity in the hippocampus. Computationally, we showed that this plasticity rule helps navigation in dynamic environments, where rewards change locations. Here, we further investigate the effects of our plasticity rule on navigation. We find that acetylcholine alters the way the simulated agent explores the environment and allows to switch between learning and unlearning. We also study how the effects of our rule vary with different characteristics of the model and the task. Finally, we show that our plasticity rule surpasses other reward-modulated plasticity rules in dynamic environments.   \\

\textbf {{\fontsize{14pt}{2pt}\selectfont Introduction  \newline}}

In order to survive, animals have to learn to interact with their environment in an effective way. 
They need to acquire new information, integrate feedback and modify their behaviour accordingly. Neuromodulation is thought to play an essential role in this process: it correlates with behavioural changes and can provide feedback and information about the environment. Dopamine, for example, acts as a reward signal and a positive reinforcement of behaviour \cite{Schultz1997, Schultz1997b, Montague1996}. Acetylcholine, on the other hand, correlates with attention \cite{Inglis1994, Inglis1995}, exploration \cite{Giovannini2001, Ceccarelli1999, Thiel1998, Acquas1996} and spatial learning in general \cite{Stancampiano1999, Fadda2000, Ragozzino1996, Ragozzino1998, Giovannini2001, Kametani1990}. \\

While the role of neuromodulation has been widely studied in the context of decision making \cite{Dayan2012}, it is still unclear exactly what neural mechanisms mediate these changes in behaviour. Research is increasingly focusing on elucidating the effects of neuromodulation on Spike-Timing-Dependent Plasticity (STDP), for which the order and precise spike timing determine sign and magnitude of plasticity. Experimentally, dopamine has been found to increase the window for potentiation in STDP \cite{Zhang2009, Edelmann2011, Yang2014}. In particular, we found that dopamine can potentiate hippocampal synapses that were previously active, even when applied after a delay \cite{Brzosko2015, Brzosko2017} . This supports the concept of an eligibility trace, which has been theorized and employed in computational modelling. An eligibility trace associates actions, and the underlying patterns of neural activity, to distal rewards \cite{Izhikevich2007, Legenstein2008, Pan2005, Suri1999, Florian2007}. Although acetylcholine has been shown to modulate synaptic plasticity in both directions \cite{Teles-GriloRuivo2013}, we found that acetylcholine biased hippocampal STDP towards depression. Interestingly, this effect could be retroactively converted into potentation by consequent application of dopamine \cite{Brzosko2017}. \\

Building on this experimental evidence \cite{Brzosko2015, Brzosko2017}, we have previously investigated the possible functional effects of sequentially neuromodulated plasticity (sn-Plast) \cite{Brzosko2017}. Using a bottom-up approach, we incorporated this novel rule into a spiking neural network model of reward-driven navigation \cite{Foster2000, Vasilaki2009, Fremaux2010, Fremaux2013}. 
We found that sequential cholinergic and dopaminergic modulation of plasticity allows flexible learning, particularly useful in dynamic environments with changing reward locations \cite{Brzosko2017}. \\ 

Here, we set out to further investigate the functional roles of sn-Plast. Inspired by experimental observations of cholinergic effects on behaviour, we examine exploration and flexibile learning in particular. In order to confirm and expand on our previous findings, we compare our rule to other types of plasticity. We show how the effects of neuromodulated STDP on behaviour depend on various model features, including state and action spaces, maze geometry and task details. We then compare sn-Plast to other types of plasticity. This allows us to deepen our mechanistic understanding of the model, and shines some light on the complex relationship between synaptic and behavioural learning. \\

\textbf {{\fontsize{14pt}{2pt}\selectfont Results  \newline}}

We base this work on recent experimental results that shine light on how hippocampal plasticity is affected by neuromodulation. In particular, dopamine was shown to retroactively potentiate previously active synapses, even when applied after a delay (up to minutes) \cite{Brzosko2015}. This provides evidence for the existence of an eligibility trace, a mechanism formerly proposed in the reinforcement learning literature as a solution to the credit assignment problem. Acetylcholine, on the other end, was found to induce depression in active synapses, regardless of the precise spike order \cite{Brzosko2017}. \\

Based on these experimental findings, we propose a spike-timing dependent plasticity rule (Fig 1A.i). We then  explore the functional roles of our neuromodulated learning rule in a neural network (Fig 1A.ii). Given the established role of dopamine as a reward signal and the increased release of acetylcholine during exploratory behaviours, we model navigation, specifically a task where the agent has to learn the path to the reward \cite{Fremaux2013}. \\

\textbf {{\fontsize{12pt}{2pt}\selectfont Cholinergic depression yields systematic exploration \newline}}

\textbf {{\fontsize{12pt}{2pt}\selectfont Radial arm maze - discrete model.}} We start our investigation with a simplified network model of a radial arm maze test (Figs 1B.i-ii). At the beginning of each trial, the agent is positioned at the centre of the maze. From there, it has to decide to which of the eight arms to move. One of the arms contains a reward (e.g. upper-central arm in Figs 1B.i-ii), which the agent has to find and learn  to reach. The network model (Fig 1A.i) of this task is composed by: i) a single presynaptic neuron, which can be thought of as a \textit{place cell} coding for the position of the agent in the maze; and ii) a post-synaptic layer of eight neurons, each representing a different arm. For clarity, we call these \textit{action neurons}, since they represent the action to take from the current position. When the trial starts and the agent is positioned in the centre of the maze, the place cell starts spiking (an inhomogeneous Poisson process, where the rate depends on the position of the agent). This, in turn, excites the action neurons (SRM$_0$, \cite{Gerstner2002}). Due to a winner take-all connectivity in the post-synaptic layer, one of the neurons is always significantly more active by the end of the trial. The winner determines the arm that will be chosen (see Methods). \\

We implement our plasticity rule on the feed-forward connections between the place cell and the action neurons. Whenever the agent finds the reward, dopamine is released globally at all synaptic sites. Thanks to the eligibility trace, only the synapses most active during the trial get retroactively potentiated. The agent can therefore successfully learn which action leads to the reward (Fig 1B.v) \cite{Izhikevich2007, Legenstein2008, Florian2007}. 
In addition, we assume acetylcholine to be present during exploration \cite{Stancampiano1999, Fadda2000, Ragozzino1996, Ragozzino1998, Giovannini2001, Kametani1990}, but not consummatory behaviour \cite{Marrosu1995, Inglis1994}. This effectively translates into the active synapse being depressed by the end of the trial, as shown experimentally in \cite{Brzosko2017}. Thus, once an arm has been chosen, there are two possible outcomes: i) the arm is rewarded, dopamine is delivered and synaptic depression is converted to potentiation, ii) the arm is not rewarded and the synapse remains depressed. \\ 

Whereas dopamine is essential to learn from a reward, acetylcholine allows learning from negative outcomes \cite{Brzosko2017}. The agent is able to exclude the unrewarding arms it has already tried from future options, thereby achieving what we call \textit{systematic exploration}. If the effect of acetylcholine is not included in the model (-ACh), the initial exploration of the maze is entirely random. The first successful trial is thus a random variable that follows a geometric distribution with $p=\frac{1}{8}$ (Fig 1B.iii) and mean 8. If, on the other hand, we assume perfect systematic exploration, an agent has a probability $\frac{1}{8}$ of finding the reward in the first trial, $\frac{7}{8}\frac{1}{7} = \frac{1}{8}$ in the second trial, $\frac{7}{8}\frac{6}{7}\frac{1}{6} = \frac{1}{8}$ in the third trial and so forth.  The first rewarded trial is distributed as a discrete uniform random variable over the interval $[1,8]$ (Fig 1B.iv, filled circles). Numerical simulations of agents with cholinergic depression (+ACh) appear to closely match the theoretical distribution (Fig 1B.iv, histogram) which means that the agent never takes more than 8 trials to find the reward.  
Systematic exploration leads to the reward faster, thus enhancing the overall performance (Fig 1B.v). Systematic exploration is further shown in a different test experiment, where we use an identical but completely unrewarded task. In this case, all the +ACh agents simulated manage to fully explore the environment at trial 8, whereas approximately half of the -ACh agents need more than 20 trials to visit all arms ($M=10000$ simulations, Fig 1B.vi). \\

 \textbf {{\fontsize{12pt}{2pt}\selectfont T-maze - continuous model.}} The radial arm maze example is a very simple model of navigation that could be practically reduced to a single decision making problem. We therefore move to a more detailed, but similar model (Methods). The basic structure of the network was kept unchanged, but new features were introduced, in particular: i) infinite possible positions for the agents (inside of the maze), ii) infinite possible actions, and iii) online decision-making at each timestep. \\

In this task, the maze has the shape of a T (Fig 1C.i-ii). There are $N=441$ place cells distributed as a grid along the stem and the arms. The position of the agent is represented by a vector of its Cartesian coordinates and can be anywhere in the maze (a continuous state space). Action neurons represent each a different direction, and are arranged in a winner-take-all fashion, as before. Decisions are taken at every timestep. The direction and speed of each move is taken to be the average of the action neurons' directions, weighted by their firing rate. This means that any arbitrary direction can be chosen (continuous action space). The recurrent connectivity ensures that only neurons with similar orientations are active at the same time. This coherent bump of activity  creates smooth and consistent trajectories. The agent is limited inside the maze: if it tries to cross the boundaries, it instantly bounces back in the opposite direction. \\

In order to test the effects of acetylcholine on systematic exploration in this model, we use a task similar to the radial maze. Each trial starts with the agent in the stem and ends when the agent enters one of the arms (or when a time limit has passed). A reward is placed in one of the arms (e.g. right arm in Fig 1C.i-ii). The agent has to discover it and learn how to reach it. \\

While cholinergic depression still makes the discovery of the reward faster (Fig 1C.i-iv) and achieves a better performance on average (Fig 1C.v $M=1000$ simulations), the empirical distribution does not match the theoretical distribution perfectly ($\sim $ Uniform[1,2]; Fig 1C.iv, +ACh). When the reward is removed, full exploration of the environment is still aided by cholinergic depression. The agent fully explores the maze in just two trials in about $85\%$ of the simulations (Figure 1C.vi). However, exploration of the environment is more systematic with acetylcholine, but not perfect. A clearcut exclusion of wrong choices was easier to obtain in the radial maze, where there was a one-to-one correspondence between the arm and the synapse. Here in the T-maze, which has more decision points and continuous actions, more complex dynamics come into play. It is still possible to suppress wrong actions (mainly because the geometry of the maze translates into a sort of discretization of the action space), but it is difficult to achieve the same level of precision. This concept will become clearer in the following sections, where we investigate this mechanism further by changing the maze to an open field. In an open field, no discretization is possible. \\

 \textbf {{\fontsize{12pt}{2pt}\selectfont Cholinergic depression enhances exploration of the action space \newline}}

  \textbf {{\fontsize{12pt}{2pt}\selectfont Learning in an open field.}} The influence of cholinergic depression on systematic exploration seems to be more complex in the continuous model. In order to study this in more detail, we choose an environment where the agent can move more freely, the open field. We model the field as a square, with place cells evenly distributed over the entire area. This task is analogous to the previous ones: the agent starts each trial in the centre of the square, and has to find and learn how to reach the reward location (circle in the top right corner of the field; Fig 2A.i-ii). \\ 

Again, due to the retroactive effect of dopamine, the vast majority of the agents are able to learn the task and navigate to the reward (Fig 2A.iv; M=1000 simulations under each condition) increasingly faster (Fig 2A.v, time to reward). Nevertheless, unlike previous tasks, cholinergic depression provides here only a marginal improvement in performance (Fig 2A.iv), and no advantage in reward discovery (Fig 2A.iii, reward discovery). Thus, in this particular environment, cholinergic depression does not seem to affect systematic exploration. \\

  \textbf {{\fontsize{12pt}{2pt}\selectfont Exploration in an open field.}} We therefore decide to further investigate the exploration patterns of our models. We remove the reward, and let the agent explore. As a proxy measure of the patterns of exploration over the field, we take the place cells' mean firing rates (average across time and simulations; Figs 2B.i-iv). Once normalized to 1, place cells' activity can be thought of as a probability distribution over the open field. This provides us with a proxy for establishing where in the field the average -ACh and +ACh agents spend the longest time. Whilst the average +ACh agent spends more time around the centre of the field (starting position; Fig 2B.ii), the average -ACh agent spends relatively more time at the boundaries (Fig 2B.i). In order to quantify the amount of exploration, we compare the distribution under the two conditions (+ACh and -ACh) with a benchmark distribution, using the Kullback-Leibler divergence (KL) as a metric. The benchmark distribution shows place cells' activity during perfectly random exploration of the environment (Fig 2B.iii). It was calculated by sampling ($M=1000$) random locations inside the open field for the duration of a trial. The average agent explores the environment more evenly without aceylcholine interaction (-ACh: KL=0.03;+ACh: KL=0.07). However, averaging over all simulations provides only limited information about the behaviour of the single realization (as an extreme example, it could happen that each agent explores only one of the corners for the entire duration of the trial, but that it chooses one of the four corners with equal probability). We therefore calculate the KL divergence between the output of each simulation and the benchmark (Fig 2B.v). According to this analysis, acetylcholine seems to modestly enhance exploration. The reason for this discrepancy is that, without cholinergic depression, there is no way of suppressing unrewarding choices. Nothing prevents -ACh agents from spending a long time in the same area at the boundaries, whereas cholinergic depression encourages +ACh agents to change direction. As a consequence, -ACh agents bounce against the walls of the maze significantly more often than +ACh agents (Fig 2B.vi). \\ 

The analysis above does not definitively answer our question about exploration. On the contrary, it further highlights the complexity of our enquiry. Exploration of the environment is affected by many factors, such as the probability of getting stuck against the walls or the precise geometry of the maze. This is because, by characterizing exploration over the open field, we are not really capturing the raw effect of acetylcholine, but rather observing its indirect consequences. In fact, by depressing synapses every timestep of the trial, acetylcholine essentially downregulates the input to the action neurons, which in turn makes the activity bump of the postsynaptic neurons move more often. The difference between consecutive actions is indeed much greater in the presence of acetylcholine (Fig 2B.vii). This translates into more circular trajectories that tend to be focused around the centre of the maze rather than the boundaries (Fig 2B.ii). \\

We can understand these findings by considering that cholinergic depression makes the activity bump more mobile, enhancing exploration over the action space. This, however, does not necessarily translate into increased exploration over the environment. In order to confirm this we perform more benchmark simulations ($M=1000$), this time as a proxy for perfect exploration over the action space. The position of the agent in each simulation is initialized at the centre of the field. From there, every action is taken completely at random: the angle of the direction is chosen from a uniform distribution over $[0, 2\pi]$, while the velocity is kept fixed (Methods). 
Place cells' activity shows a high peak around the initial position (Fig 2B.iv), meaning that the average benchmark agent does not move very far. As expected, the distribution of place cells' activity in the +ACh simulations (Fig 2B.ii) is more similar to this benchmark distribution than -ACh simulations (-ACh: KL =13.12, Fig 2B.i; +ACh: KL = 9.7, Figure2B.ii). \\

 \textbf {{\fontsize{12pt}{2pt}\selectfont Cholinergic depression improves performance in dynamic environments \newline}}

  \textbf {{\fontsize{12pt}{2pt}\selectfont Relearning in an open field.}} We have shown that cholinergic depression allows the agent to learn from negative outcomes and increases exploration over the action space. These characteristics suggest that cholinergic depression might be especially advantageous in dynamic environments. We consider a task in which, after an initial 20 trials where the agent learns how to navigate to the reward (Fig 2A), the reward is moved to a new location. In our case, it is moved to the opposite corner (Figs 3A.i-ii). +ACh agents  discover the new reward location in fewer trials, while as much as one out of four -ACh agents cannot find it before the end of the experiment (Fig 3A.iii) \cite{Brzosko2017}. In addition, the +ACh agents show better task performance than the -ACh agents   ($96.8\%$ correct versus $63\%$ correct; Fig 3A.v). -ACh agents  mostly just extend the previously learned path (Fig 3A.ii), whereas +ACh agents are able to unlearn the old reward location (Figs 3A.i, 3A.iv). This results in a difference in the time to navigate to the new reward location (Fig 3A.vi). \\

When we consider the average place cells' activity in the different trials (Fig 3B), we note that cholinergic depression allows unlearning of the previously rewarded location. This increases exploration (trial 26, place cells in the field are more uniformly active) and allows the agent to eventually learn the path to the new reward (trial 31). In the absence of acetylcholine, place cells' firing rates peak around the first reward location for the entire duration of the experiment, with only a marginal increase of activity at the newly rewarded area (trial 31). The patterns of activity at trial 21 show that acetylcholine leads to a more precise trajectory. This is because an agent equipped with cholinergic depression can forget those actions that do not lead to the reward. For example, if a +ACh agent takes a path that arrives close to the reward but misses it, the corresponding synapses are promptly depressed and therefore forgotten. Without acetylcholine, on the other hand, there is a slow broadening of the optimal path and a loss of precision. This is also reflected in the difference in performance (Fig 2A.iv). \\

  \textbf {{\fontsize{12pt}{2pt}\selectfont Learning and relearning in an open field with obstacles.}} As mentioned earlier, the specifics of the task strongly affects the outcome. To explore this point further, we repeat the same experiment using a slightly different maze geometry. We insert two vertical obstacles in the open field, and move the reward location on the x axis, to the right side of the obstacles, for the first part of the experiment (Figs 4A.i-ii; obstacles = white vertical bars, reward location = black solid circle). In this case, -ACh agents initially perform better at finding the reward (Fig 4A.iii). It is much easier to discover the reward when following straight lines in this particular maze geometry (even more so than in a simple open field; Figs 4A.i-ii). Later in the experiment (Trial 20), however, agents equipped with cholinergic depression achieve a slightly higher success rate (Fig 4A.iv), and are faster to navigate to the reward (because they do not get stuck against the walls or the obstacles; Figs 4A.i-ii,v). The results for  the second part of the experiment, when the reward is moved horizontally to the left side of the obstacles, are qualitatively similar to the open field but even more pronounced (Fig 4B). Almost $40\%$ of -ACh agents ($39.7\%$) do not find the new reward before the end of the experiment (Fig 4B.iii), and $89.2\%$ of them still visit the old reward location in the last trial (Fig 4B.v). With this maze geometry, it is more difficult to extend the old path to the new reward location. More prominently than in the open field, agents with cholinergic depression are twice as successful as -ACh agents (Fig 4B.v) and can navigate to the reward twice as fast (Fig 4B.vi). \\

 \textbf {{\fontsize{12pt}{2pt}\selectfont Comparison with other learning rules \newline}}

 \textbf {{\fontsize{12pt}{2pt}\selectfont Reward-modulated STDP.}} Until now, we have focused on the functional role of cholinergic depression, comparing the same learning rule with and without cholinergic depression (+ACh and -ACh). We next investigate how sn-Plast compares to other reward-modulated learning rules. \\

First, we change the plasticity rule in our model to standard reward-modulated STDP (r-STDP; Fig 5). In r-STDP, synapses follow a classical STDP rule with different amplitudes for the pre-post ($A_{pre-post}$) and post-pre ($A_{post-pre}$) windows. However, all synaptic changes are gated by dopamine and become effective only retroactively through an eligibility trace. When  $A_{pre-post} = A_{post-pre} = +1$, reward-modulated STDP is equivalent to the plasticity rule used in our control simulations (sn-Plast without acetylcholine; -ACh). \\ 

In order to test how the shape of the STDP window affects learning, we perform a parameter sweep. The task is the same: the agent moves in an open field and has 20 trials to learn to navigate to the reward location (Fig 2A). We  examine how the mean percentage of successful trials ($M=200$ simulations; Fig 5A.i) varies with the amplitudes of the learning window ($A_{pre-post}$ and $A_{post-pre}$). When there is no learning  ($A_{pre-post}=A_{post-pre}=0$; middle square, Fig 5A.i), the average performance is $37\%$. Using this value as baseline, we can determine whether agents learn or unlearn. The agents' performance varies with the integral of the learning window: it rises above baseline for positive-integral windows ($A_{pre-post}+A_{post-pre}>0$; the part above the diagonal, Fig 5A.i) and below baseline for negative-integral windows ($A_{pre-post}+A_{post-pre}<0$; the part under the diagonal, Fig 5A.i). When the integral of the plasticity window is zero  ($A_{pre-post}+A_{post-pre}=0$; diagonal of the matrix, Fig 5A.i), there is little variation from baseline. However,the performance clearly increases with the amplitude of the pre-post learning window, $A_{pre-post}$ (Fig 5A.ii). This is because, in a  spiking neural network, presynaptic spikes contribute to elicit postsynaptic spikes (spike-spike correlation, \cite{Kempter1999}). As such, the amplitude of the pre-post window, relatively to the post-pre window, is partly involved in learning. We can conclude that the order of the spikes matters, although only marginally so. In our model, what really determines whether the agent learns or unlearns is the integral of the STDP window. \\

We next compare four agents, equipped with different STDP windows having: i) positive integral (red), ii) negative integral (yellow), iii) zero integral and $A_{pre-post} > A_{post-pre}$ (dark orange) and iv) zero integral and $A_{pre-post} < A_{post-pre}$ (inverse STDP window; light orange). As expected, the best learner is the agent with  the positive learning window, whereas the agent with a negative learning window effectively unlearns (Fig 5B.ii). There is generally very little change in performance when the integral of the STDP window is zero: if $A_{pre-post} > A_{post-pre}$, we can observe some slow learning; if $A_{pre-post} < A_{post-pre}$ there is very slow unlearning instead (Fig 5B.i). As mentioned earlier (Fig 5A.ii), the spike order is still relevant, although marginally so. This is due to spike-spike correlation \cite{Kempter1999}. These patterns remain consistent in the second part of the experiment, when the reward is moved to a different corner of the field. The agent with a positive integral learns how to navigate to the new reward location (Fig 5B.iv) but does not really unlearn the path to the first reward (the visits to the previously rewarded location are still as high as $62.4\%$ at trial 40; Fig 5B.iii). The agent with a negative integral completely unlearns the path to the second reward too (Fig 5B.iv). Agents with vanishing integrals show very little change in both learning of the new reward location and unlearning of the old one (Figs 5B.iii-iv). \\

Thus, r-STDP allows the agent to either learn or unlearn the path to the reward, depending on the integral of the learning window. In contrast, our sequentially neuromodulated plasticity rule can flexibly switch between these modalities in response to environmental changes. For this reason, sn-Plast is more suited to dynamic tasks that require a degree of adaptation. This analysis also shows how the spike order is only relatively important to learning. This characteristic is intrinsic to the model \cite{Fremaux2010, Fremaux2013} and in striking agreement with the experimental data from which we derive our plasticity rule (both dopaminergic and cholinergic modulated STDP windows are symmetric and therefore invariant to spike order; \cite{Brzosko2015, Brzosko2017}). \\


 \textbf {{\fontsize{12pt}{2pt}\selectfont Negative feedback.}} In our model, we assume that synaptic plasticity is biased towards depression unless a reward is delivered. Even if it helps to unlearn a previously learned sequence of actions, this kind of depression is indiscriminately persistent throughout exploration and is not specific to reward omission. Alternatively, we could imagine that a negative feedback is delivered to the synapse when the expected reward is omitted. The sign of the synaptic change would then be positive if the reward is delivered ($A_{feedback}=1$) and negative if it is omitted ($A_{feedback}=-1$). This feedback signal is reminiscent of a prediction error, but different in that the expectations are not updated during the experiment \cite{Schultz1997}. We thus compare sn-Plast to this model with targeted negative feedback. As before, the agent explores the open field  for the first 20 trials and has to learn how to navigate to the reward. For the remaining 20 trials the reward is moved to the opposite corner of the field, and the agent has to discover it and learn the new path. Unlike before, however, a trial ends if the agent enters either one of the rewarding areas (old or new), or if a time limit is reached. Whenever the agent enters the old reward location, a negative feedback signal induces synaptic depression. \\

Since no negative feedback is present in the first half of the task, agents with negative feedback signal perform identically to -ACh agents (Fig 6A). Agents with continuous updating of cholinergic depression increases the success rate (Fig 6A.ii) and diminishes the average time to reward (Fig 6A.iii), but it slightly deteriorates the initial exploration (it takes longer to discover the reward; Fig 6A.i). It is in the second part of the experiment, after reward displacement, that we see the effect of the negative feedback. As expected, -ACh agents show the poorest performance (Fig 6B). In contrast, agents with negative feedback signal are able to partially unlearn the previously rewarded location (Fig 6B.ii). They can therefore also find (Fig 6B.i) and reach (Fig 6B.iii) the newly rewarded location more often.  Nevertheless, +ACh agents show the best results. They unlearn the old reward location completely (Fig 6B.ii). They also find and learn the new reward location, reaching an almost perfect performance (Fig 6B.iii). The time to the reward is also shorter for +ACh agents, whereas almost no difference can be found between the other two sets of simulations (Fig 6B.iv). \\

Targeted negative feedback allows unlearning of previously rewarded actions to some extent. However, it performs quite poorly when compared to cholinergic depression, at least in the current model. Online cholinergic depression seems to offer a more powerful and direct way of forgetting. \\ 

\textbf {{\fontsize{14pt}{2pt}\selectfont Discussion  \newline}}

In this paper we investigated the possible functional consequences of neuromodulated hippocampal STDP, based on our recent experimental findings \cite{Brzosko2017}. In particular, we analyzed this plasticity rule in a network model of reward-driven navigation. Consistent with previous models, dopamine makes it possible to learn the path to the reward \cite{Izhikevich2007, Legenstein2008, Florian2007}. Acetylcholine, instead, allows learning from negative outcomes. This yields behavioural flexibility and is particularly useful in dynamic environments, where it is necessary to both learn and unlearn in a task-relevant manner. In a simple model with discrete state and action space, cholinergic depression allows suppression of unrewarding choices and systematic exploration of the maze. In more complex continuous models, it enhances exploration over the action space, but this does not necessarily translate into increased exploration over the entire maze.  \\

 \textbf {{\fontsize{12pt}{2pt}\selectfont Dopamine.}} Dopamine is thought to signal reward delivery and reinforce behaviour \cite{Schultz1997, Schultz1997b, Montague1996}. The behavioural \cite{rescorla, Pavlov1927, Thorndike1911} and algorithmic \cite{Sutton1998} mechanisms of reward-modulated learning have been thoroughly investigated and characterized. Recently, its neural substrates have been explored as well. Dopamine has been reported to shift STDP towards potentiation \cite{Zhang2009, Edelmann2011, Yang2014}. In this study, we build on our previous experimental findings on the retroactive effect of dopamine on hippocampal synaptic plasticity  \cite{Brzosko2015}. Taking inspiration from both reinforcement learning \cite{Sutton1998} and biology \cite{Lisman1989}, dopamine was theorized to act on synaptic eligibility traces. These traces keep track of and reinforce neural activity associated with distal rewards \cite{Izhikevich2007, Legenstein2008, Florian2007}. Given the similarity of sn-Plast to other reward-modulated plasticity rules \cite{Izhikevich2007}, our network unsurprisingly succeeds in learning rewarded patterns of activity. Nevertheless, our learning rule does differentiate itself from other reward-modulated plasticity rules because of the symmetrical positive shape of its STDP window. Consistently, in our model the integral of the learning window has greater importance than the exact spike timing \cite{Kempter1999}. In a different framework, however, spike timing could have functional roles, for instance when precise spike sequences are learned \cite{Izhikevich2007, Legenstein2008, Fremaux2010}. \\

One limitation of our model is that we assume dopamine to signal reward exclusively. As such, it would only update synaptic weights during reward delivery. However, dopamine has also been associated with spatial novelty \cite{McNamara2014, Lisman2005, Li2003, Otmakhova2013, Tran2008, Ihalainen1999}, and roles for dopamine beyond goal-directed navigation are quite likely \cite{de2015explicit, Atherton2015}. \\ 

 We approximate dopamine as a stable reward signal that is available globally at the synapses with every reward delivery. However, dopaminergic neurons exhibit different modes of firing, with phasic firing coding for reward prediction error. As such, dopamine is released only when the reward is unexpected \cite{Montague1996, Schultz1997,  Schultz1997b}. If the animal is able to predict the reward delivery correctly, then VTA dopaminergic neurons would not increase their firing rate. It would be interesting to see how including a predicion error signal in our model would affect the results.  \\

\textbf {{\fontsize{12pt}{2pt}\selectfont Acetylcholine.}} Acetylcholine is known to play an important role in learning and memory \cite{Hasselmo2006, Deiana2011, Easton2012}. In the hippocampus, acetylcholine has been reported to facilitate both long term potentiation \cite{Boddeke1992, Huerta1995, Ovsepian2004,Shinoe2005,Buchanan2010,Connor2012,  Digby2012, Dennis2016, Adams2004, Sugisaki2011, Sugisaki2016} and long-term depression \cite{Scheiderer2006, Volk2007, Dickinson2009, Jo2010, Kamsler2010}, depending on a number of variables, such as plasticity induction protocols, acetylcholine concentrations and type of cholinergic receptors \cite{Teles-GriloRuivo2013}. In our previous work \cite{Brzosko2017}, we found that cholinergic modulation of hippocampal STDP resulted in a symmetrical negative learning window and used this data as a starting point for our investigation.  \\

Acetylcholine has been studied in relation to behavioural tasks \cite{Picciotto2012, Pepeu2004, Deiana2011}. Microdialysis studies have reported an increase in cholinergic release in the hippocampus during engagement in spatial learning tasks \cite{Stancampiano1999, Fadda2000, Ragozzino1996, Ragozzino1998, Thiel1998, Giovannini2001, Kametani1990} and a reduction during consummatory behaviour \cite{Inglis1994, Marrosu1995}. Adding this dynamic to our neuromodulated network allowed us to study the possible effects of acetylcholine on navigation and decision-making \cite{Brzosko2017}. Acetylcholine has been postulated to signal novelty and saliency \cite{Giovannini2001, Ceccarelli1999, Acquas1996}, and was reported to enhance exploratory behaviours like rearing \cite{Thiel1998, Flicker1982}. For this reason, we largely focused on characterizing the effect of acetylcholine on exploration. In our model, acetylcholine indeed increases exploration over the action space, but this does not necessarily translate into increased exploration over the entire physical environment. In addition to exploration, the effect of acetylcholine on spatial learning has also been connected to paradigm shifts and reversal learning \cite{Tzavos2004, McCool2008, Ragozzino2002}, which in turn has been shown to depend on long-term depression in hippocampal synapses \cite{Dong2013}. This is in agreement with our observation that acetylcholine is useful in dynamic scenarios, where forgetting previously learned actions is advantageous. \\

Acetylcholine has been hypothesized to modulate learning in other computational theories before \cite{Doya2002, Yu2005, Hasselmo1999}. It was put forward as a signal for uncertainty in probabilistic environments \cite{Yu2005} and a switch signal for the encoding of new information, as opposed to the consolidation of memories \cite{Hasselmo1999, Doya2002}. Finding a clear correspondence between these theories and the model we present here is not trivial. However, our results are consistent with previous work, in that they suggest a functional role for acetylcholine in learning which is: i) complementary to dopamine, and ii) relevant to dynamical, changing environments. \\

\textbf {{\fontsize{12pt}{2pt}\selectfont Conclusion.}} In conclusion, we model here a role for dopamine as a behavioural reinforcer, and propose a new role for cholinergic depression in learning from negative outcomes. Despite its simplicity, our feed-forward network captures the key characteristics of sequentially neuromodulated plasticity, allowing us to examine its potential role in reward-based  navigation \cite{Brzosko2017}. In addition, by allowing us to clearly examine its dynamics, it provides us with a useful tool to further investigate the relationship between synaptic and behavioural learning. The continuosly updated cholinergic depression allows learning from unsuccessful trials, unlearning of previous reward locations, and enhanced exploration over the appropriate action space. As such, sn-Plast is an effective reward-modulated learning rule for navigation tasks. \\

\textbf {{\fontsize{14pt}{2pt}\selectfont Methods  \newline}}

The navigation model is based on a one-layer network \cite{Fremaux2013}. The \emph{place cells} in the input layer code for the position of the agent in the environment. They project to the output layer of \emph{action neurons}. Each one of the action neurons represents a different direction. Lateral connectivity in this layer ensures that action neurons compete with each other in a winner-take-all scheme. Their activity is then used to determine the action (i.e. direction and velocity) to take at every instant. \\
\\

\textbf {{\fontsize{12pt}{2pt}\selectfont Place cells \newline}}

\textbf {{\fontsize{12pt}{2pt}\selectfont Discrete model.}} In the case of the radial maze, the state space is discrete and contains only one location: the centre of the maze. From there, the agent chooses to which of the eight possible arms to move. The network is therefore composed of a single place cell, active for the whole duration of the trial, simulated as a Poisson process with rate $\bar{\lambda}^{pc} = 4000$ Hz. \\
\textbf {{\fontsize{12pt}{2pt}\selectfont Continuous model.}} The position of the agent at time t is described by the bi-dimensional vector of its Cartesian coordinates, $\textbf{x}(t)$. There are $N$ place cells, spread over the entire environment at a horizontal and vertical distance of $\sigma$ from one another. The spiking activity of place cell $i$ is modelled as an inhomogeneous Poisson process, with rate $\lambda^{pc}_i(\textbf{x}(t))$ defined as follows: $$\lambda^{pc}_i(\textbf{x}(t)) = \bar {\lambda}^{pc} \exp \Big( -\frac{||\textbf{x}(t) - \textbf{x}_{i}||^2}{\sigma^2}\Big).$$
The firing rate $\lambda^{pc}_i$  is a function of the distance of the agent from the place cell centre $\textbf{x}_{i}$. It is at its maximum, $\bar{\lambda}^{pc} = 400$ Hz, when the agent is located exactly in $\textbf{x}_{i}$ and it decreases as it moves away. This mechanism simulates a place field in a 2D environment, which allows for an accurate representation of the position of the agent in the environment. \\
\textbf {{\fontsize{12pt}{2pt}\selectfont Open field - continuous model.}}The open field is modelled as a square of side length of 4 a.u. The initial position of the agent in each trial is the centre of the open field, which corresponds to the origin of the Cartesian plane.When obstacles are added, they are modelled as two rectangular bars of sides $sx_{obs}=0.4$ a.u. and $sy_{obs}=0.8$ a.u. centred on the x axis at $x_1=-1.2$ a.u. and $x_2=1.2$ a.u.. In the open field, there are $N=221$ place cells at distance $\sigma=0.4$ from one another. \\
\textbf {{\fontsize{12pt}{2pt}\selectfont T-maze - continuous model.}}The T-maze is cropped out from the open field plane. It is composed by a stem of length $l_{stem}=3.2$ a.u. and width $wd_{stem}=0.6$ a.u., and two arms, each having length $l_{arm}=1.7$ a.u. and width $wd_{arm}=0.8$ a.u.. The agent starts every trial from the bottom of the stem: $(x_{start}, y_{start})=(0,-2)$ a.u.. In the T-maze, there are $N=442$ place cells at distance $\sigma=0.2$ from one another. \\
\\

\textbf {{\fontsize{12pt}{2pt}\selectfont Action neurons  \newline}}

\textbf {{\fontsize{12pt}{2pt}\selectfont Neuron model.}} Place cells constitute the input to the network, and they all project to all action neurons with weights $w^{feed}$. These feed-forward weights are initialized to $w_{in}$ and bounded between $w_{min}$ and $w_{max}$ (see Table 1 for specific values). Action neurons are also connected with each other through synaptic weights $w^{lat}$. The neurons are modelled as SRM$_0$ (\cite{Gerstner2002}), the membrane potential of neuron $j$ is therefore given by:
$$u_j(t) =\sum_i \sum_{\bar t_i \in F_i^{pc}, t>\hat{t}_j} w^{feed}_{ji} \cdot \epsilon(t-\bar t_i) + \sum_{k, k \ne j} \sum_{\bar t_k \in F^{a}_k, t>\hat{t}_j} w^{lat}_{jk} \cdot \epsilon(t-\bar t_k)+ \chi \Theta(t-\hat{t}_j)\exp \big( -\frac{t-\hat{t}_j}{\tau_m} \big), $$
where $\chi=-5$ mV scales the refractory period, $\hat{t}_j$ is the last postsynaptic spiking time and $\epsilon$ is the EPSP described by the kernel $\epsilon(t) = \frac{\epsilon_0}{\tau_m-\tau_s} \Big( e^{\frac{-t}{\tau_m}}- e^{\frac{-t}{\tau_s}}\Big) \Theta(t), $ with $ \Theta(t)$ being the Heaviside step function, $\tau_m=20$ ms, $\tau_s=5$ ms, $\epsilon_0=20$. $ F^{pc}_i$ and $F^{a}_k$ are sets containing respectively $\bar t_i$ and $ \bar t_k$ , the arrival times of all spikes fired by place cell $i$ and action neuron $k$. Spiking behaviour is stochastic and follows an inhomogeneous Poisson process with parameter $\lambda_j(u_j(t))$, which depends on the membrane potential at time $t$. In particular, $$\lambda_j(u_j(t))= \lambda_0 \exp \Big(\frac{u_j(t) - \theta}{\Delta u} \Big),$$
where $\lambda_0$ is the maximum firing rate, $\Delta u$ regulates randomness of the spiking behaviour and $\theta=16$ mV is the spiking threshold. \\
\begin{table}[h!]
\caption{}
\centering
\begin{tabular} {c rrrr}
\hline \hline
& Radial arm & T-maze & Open Field \\
\hline 
$w_{min}$ &1 &1 &1 \\
$w_{max}$ &5 &5 &3 \\
$w_{in}$ &2 &3 &2 \\
$\eta_{\text{ACh}}$ &0.001 &0.1 &0.002 \\
$\eta_{\text{DA}}$ &0.01 &0.3 &0.01 \\ 
\hline
\end{tabular}
\end{table}

\textbf {{\fontsize{12pt}{2pt}\selectfont Discrete model.}} In the radial maze, there are only eight possible actions to take from the initial position. There are $N=8$ neurons, each coding for a different arm. These neurons are connected through inhibitory synapses: $w^{lat}=-250$. This connectivity scheme ensures that, given enough time, one neuron will inhibit all others and be significantly more active. Other parameters were set to: $\lambda_0=100$ Hz, $\Delta u=0.5$ mV.\\
\textbf {{\fontsize{12pt}{2pt}\selectfont Continuous model.}} Action neurons represent different directions in the Cartesian plane. Specifically, each action neuron $j$ represents direction $\textbf{a}_j$, where $\textbf{a}_j=a_0(\sin(\theta_j), \cos(\theta_j))$, with $\theta_j = \frac{2j\pi}{N}$, $N=40$ and $a_0=0.08$. The lateral connectivity between action neuron $k$ and action neuron $j$ is defined as follows $$w^{lat}_{jk} = \frac{w_{-}}{N}+w_{+}\frac{f(j,k)}{N},$$ where ${w_{-}}=-300$, ${w_{+}}=100$ and $f$ is a lateral connectivity function, which is symmetric, positive and increases monotonically with the similarity of the actions. In particular, $f(j,k) = (1-\delta_{jk}) e^{\psi \cos(\theta_j-\theta_k)}, $ with $\psi=20$. Neurons therefore excite each other when they have a similar tuning, and depress otherwise. This ensure that only a few similarly tuned action neurons are active at any given time, making the trajectory of the agent smooth and consistent. Other parameters were set to: $\lambda_0=60$ Hz, $\Delta u=2$ mV. \\
\\


\textbf {{\fontsize{12pt}{2pt}\selectfont Action selection  \newline}}

The action selection process determines the decision to take, based on the firing rates of the action neurons. The activity of action neuron $j$ is approximated by filtering spike train $Y_j$ with kernel $\gamma$:  $$\rho_j(t) = (Y_j \circ \gamma)(t),$$ where  $Y_j = \sum_{\bar{t}_j\in F^a_j}\delta(t-\bar{t}_j)$ and $ \gamma(t) = \frac{e^{\frac{-t}{\tau_{\gamma}}}- e^{\frac{-t}{\nu_{\gamma}}}}{\tau_{\gamma}-\nu_{\gamma}} \Theta(t),$ with $\tau_{\gamma}=50$ ms and $\nu_{\gamma}=20$ ms. \\
\textbf {{\fontsize{12pt}{2pt}\selectfont Discrete model.}} Decisions in the discrete case are taken only at the end of the trial. When a time limit $T_{max}=5$ s has been reached, the action neuron with maximum firing rate is selected. In the unlikely case two neurons exhibit exactly the same firing rate at the end of trial, the winning neuron is chosen at random. The agent then enters the arm associated with the winning neuron. All activity is reset before the onset of the next trial.\\
\textbf {{\fontsize{12pt}{2pt}\selectfont Continuous model.}} In the continuous case, actions are taken continuously, at every timestep $t$. The action selection process thus determines $\textbf{a}(t)$, the action to take at time $t$. If each action neuron $j$ represents direction $\textbf{a}_j$ and has an estimated firing rate $\rho_j(t)$, then the action $\textbf{a}(t)$ is the average of all the directions encoded, weighted by their respective firing rates 
$$ \textbf{a}(t) = \frac{1}{N}\sum_j{\rho_j(t)\textbf{a}_j}, $$ where $N=40$ is the total number of action neurons. This decision making mechanism allows the agent to move in any direction, making the action space effectively continuous.\\
\\

\textbf {{\fontsize{12pt}{2pt}\selectfont Navigation details}}\\ 

\textbf {{\fontsize{12pt}{2pt}\selectfont Continuous model.}} Once action  $\textbf{a}(t)$ has been determined, the update for the position of the agent is 

\begin{equation*} 
\Delta \textbf{x}(t)=  \begin{cases}
\textbf{a}(t), & \text{if \textbf{x}(t+1) within the boundaries}. \\
d \cdot \textbf{u}(\textbf{x}(t)) & \text{ otherwise}
\end{cases}
\end{equation*}

The agent therefore normally moves with instantaneous velocity $\textbf{a}(t)$. 
When the agent tries to surpass the limits of the field, it is instantly bounced back by a distance $d=0.01$. The unit vector \textbf{u}(\textbf{x}(t)) points in the direction opposite to the boundary. To avoid large boundary effects, the feed-forward weights between place cells on the boundaries and action neurons that code for a direction $\textbf{a}_j$ outside of the field are set to zero.\\ 

The agent is free to explore the environment for a maximum duration of $T_{max}$. If it finds the reward at a time $t_{rew}<T_{max}$, the trial is terminated earlier, precisely at time $t=T_{rew}+300$ ms. The extra time mimics consummatory behavior, navigation is thus paused during this interval (i.e. place cells activity is set to zero). The effect of the inter-trial interval is modelled by resetting all activity. 

\textbf {{\fontsize{12pt}{2pt}\selectfont T-maze - continuous model.}} When used in the task, the reward is located in the right arm of the maze. Specifically, we consider the reward to be found whenever the agent crosses the vertical line $x_r = 1$ a.u.. The maximum duration of a trial is $T_{max}=5$ s, but the trial ends whenever the agent enters one of the arms (whenever the agent crosses either the vertical line $x_r = 1$ or the vertical line $x_l=-1$) .  When in the stem, the available actions are restricted only to upwards movements (angle between $\theta \in [\frac{\pi}{4}, \frac{3\pi}{4}]$). When in the top part of the maze, only horizontal movements are allowed (angle between $\theta \in [-\frac{\pi}{4}, \frac{\pi}{4}] \cup [-\frac{3\pi}{4}, \frac{5\pi}{4}]$ ). 

\textbf {{\fontsize{12pt}{2pt}\selectfont Open field - continuous model.}} For the first 20 trials, the reward can be found in the circular goal area centred in $c_1 = (1.5, 1.5)$ with radius $r_1=0.3$. In trials 21 to 40, the goal area moves to centre $c_2=(-1.5, -1.5)$, but maintains the same shape and size. 
If the open field has obstacles, the agent is not allowed to cross them and is therefore pushed back, similarly to what happens with the walls. In this case, the goal area is initially centred in $c_1 = (0, 1.5)$, and then moved to $c_2=(0, -1.5)$. The maximum duration of a trial is $T_{max}=5$ s. \\


%



\textbf {{\fontsize{12pt}{2pt}\selectfont Sequentially neuromodulated plasticity (sn-Plast)}}\\ 

The synaptic weights between place cells and action neurons play a fundamental role in defining a policy for the agent. Plasticity is essential for the agent to learn to navigate the open field and is implemented in a way that follows the experimental results presented in Brzosko et al. 2015 and 2017. 
The synaptic changes combine the modified STDP rule (Fig 3) and an eligibility trace that allows for delayed updates. \\

In particular, the total weight update is:  
$$\Delta w_{ji}(t) = \eta A \Big( \Big( \sum_{ \bar{t}_i \in F^{pc}_i} {\sum_{\bar{t}_j \in F^{a}_j} {W(\bar{t}_j - \bar{t}_i)}} \Big) \circ  \psi \Big)(t),  $$
where $\eta$ is the learning rate, $A$ emulates the effect of the different neuromodulators, $W$ is the STDP window and $\psi$ is the eligibility trace. $ F^{pc}_i$ and $F^{a}_j$ are sets containing respectively $\bar{t}_i$ and $\bar{t}_j$ , the arrival times of all spikes fired by place cell $i$ and action neuron $j$. \\

The basic STDP window is 
$W(x) = e^{-\frac{|x|}{\tau}}, $
with $\tau=10$ ms. This function is always symmetric and positive, but the sign of the final weight change is determined by the neuromodulators at the synapse: 
\begin{align*} 
A =  &\begin{cases}
-1 &\text{ -DA, +ACh } \\
0 &\text{ -DA,  -ACh } \\
1 &\text{ +DA, $\pm$ACh} \\
 \end{cases} \end{align*}

Dopamine is assumed to be released simultaneously in all synapses whenever a reward is delivered. All weight changes are gated by neuromodulation ($A=0$ when all neuromodulators are absent). The learning rate $\eta$ also depends on neuromodulators (see Table 1 for specific values): 
$$\begin{aligned} 
\eta=  &\begin{cases}
\eta_{\text{ACh}} &\text{ -DA, +ACh } \\
0  &\text{ -DA,  -ACh } \\
\eta_{\text{DA}} &\text{ +DA, $\pm$ACh}. 
 \end{cases} \end{aligned}$$ 
The weight change due to STDP is convoluted with an eligibility trace $\psi$, modelled as an exponential decay $\psi(t) = e^{-\alpha\frac{t}{\tau_e}}\Theta(t)$, with $\tau_e= 2$ s and $$\begin{aligned} 
\alpha= &\begin{cases}
1 &\text{ +DA } \\
0  &\text{ -DA }. 
 \end{cases} \end{aligned}$$ The eligibility trace keeps track of the active synapses and allows for a delayed update of the synaptic strength. Variable $\alpha$ in the exponent acts as a flag and ensures that the eligibility trace is active with dopamine only ($\alpha=1$).  \\
When no interaction with acetylcholine was assumed (-ACh), the weights were potentiated only at the end of the trial, in the case that the agent found the reward ($A=1, \alpha =1$). They were left unchanged otherwise ($A=0$). If acetylcholine was present throughout the task (+ACh), the weights were updated online ($A=-1, \alpha =0$). When no reward was found before the end of the trial, weights were depressed. Otherwise, they were potentiated retroactively ($A=1, \alpha =1$). \\

\textbf {{\fontsize{12pt}{2pt}\selectfont \newline  Dopamine-modulated standard asymmetric STDP curve}} \\

We also compared our symmetric learning windows to standard asymmetric STDP curves. The total weight update with this rule is
$$\Delta w_{ji}(t) = \eta \Big( \Big( \sum_{ \bar{t}_i \in F^{pc}_i} {\sum_{\bar{t}_j \in F^{a}_j} {W_2(\bar{t}_j - \bar{t}_i)}} \Big) \circ  \psi \Big)(t),  $$
where $\eta=0.01$ is the learning rate, $W_2$ is the STDP window and $ \psi$ is the eligibility trace (as defined above). $F_i^{pc}$ and $F_j^a$ are sets containing $\bar{t}_i$ and $\bar{t}_j$ respectively, the arrival times of all spikes fired by place cell $i$ and action neuron $ j$.
The spike timing plasticity rule was implemented as follows:
\begin{align*} 
W_2(s)=  &\begin{cases}
A_{pre-post}e^{-\frac{s}{\tau}}  &\text{if } s>0 \\
\frac{1}{2}(A_{pre-post} + A_{post-pre})  &\text{if } s=0\\
A_{post-pre}e^{\frac{s}{\tau}} &\text{if } s<0 
 \end{cases} 
 \end{align*}
 
The integral of the learning window determines if the agent learns, unlearns or does not learn. We therefore considered four different parameter sets: (i) positive integral ($A_{pre-post}=1$, $A_{post-pre}=-0.5$);  (ii) negative integral ($A_{pre-post}=0.5$, $A_{post-pre}=-1$); zero integral with either (iii) positive $A_{pre-post}$ (standard STDP window; $A_{pre-post}=0.5$, $A_{post-pre}=-0.5$) or (iv) negative $A_{post-pre}$ (inverted STDP window; $A_{pre-post}=-0.5$, $A_{post-pre}=0.5$). The time constant was identical for the two sides of the window and was taken to be $\tau=10$ ms. We ran 1000 simulations for each parameter set. \\

\textbf {{\fontsize{12pt}{2pt}\selectfont Negative feedback signal}} \\

We also compared our neuromodulated learning rule to a dopamine-modulated rule with negative feedback. In this set of simulations, we assumed that whenever the agent reaches the location of an omitted reward it receives a negative feedback that inverts the sign of the learning window induced by dopamine. The weight update for simulations with negative feedback is: 
$$\Delta w_{ji}(t) = \eta A_{feedback} \Big( \Big( \sum_{ \bar{t}_i \in F^{pc}_i} {\sum_{\bar{t}_j \in F^{a}_j} {W(\bar{t}_j - \bar{t}_i)}} \Big) \circ  \psi \Big)(t),  $$ where $\eta=0.01$, $A_{feedback}=1$ when the new reward is found, $A_{feedback}=-1$ if the agent navigates to the old reward location and $A_{feedback}=0$ otherwise. \\

\textbf {{\fontsize{12pt}{2pt}\selectfont Acknowledgments}}\\ 

We would like to thank Yann Sweeney and Wilten Nicola for reviewing the manuscript.



\newpage
\includegraphics[width=\textwidth]{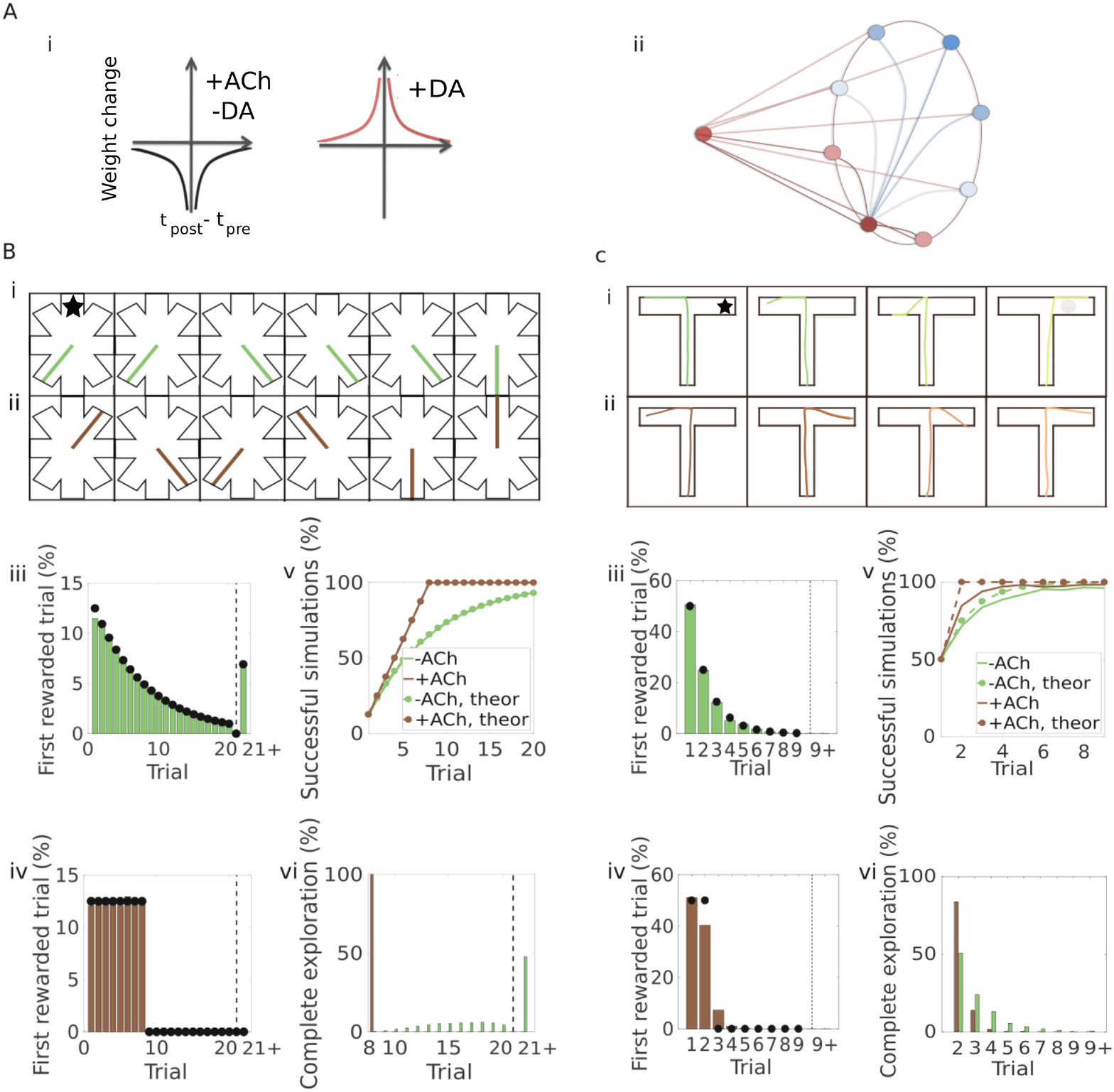} 
\textbf{Fig 1. Cholinergic depression yields systematic exploration of a radial arm maze and a T-maze.} A. i) Sequentially neuromodulated spike-timing-dependent plasticity rule. When acetylcholine is present at the synapse, the plasticity window (black) is negative and symmetric (i.e. the weight changes proportionally to the lag between the spikes, irrespectively of the order). If dopamine is added, the plasticity window converts to positive (red).  ii) Schematic of a neural network model of a reward-driven navigation task. A place cell is connected to action neurons, which inhibit each other following a winner-take-all scheme. B. \textbf{Radial maze:} i-ii) Example trajectories. The maze consist of eight arms, the reward is located in the upper-central arm (with the star). i) The agent without cholinergic depression (-ACh, green) visits the same unrewarded arm more than once. ii) The agent with cholinergic depression (+ACh, brown) explores the maze in a systematic way: it excludes unrewarded arms and finds the rewarded arm sooner. iii) Percent cumulative distribution of the first rewarded trial (histogram) and the corresponding theoretical distribution (geometric distribution with $p=\frac{1}{8}$; black dots) for simulations without cholinergic depression. iv) Percent cumulative distribution of the first rewarded trial (histogram) and the corresponding theoretical distribution (discrete uniform distribution on $[1,8]$; black dots) for simulations with cholinergic depression. v) Percentage of successful simulations over consecutive trials for -ACh and +ACh agents (solid lines). Theoretical learning curves, assuming one-shot learning (cumulative distribution of the first successful trial; dashed lines with dots). vi) Agents navigating the environment without any reward. Percent cumulative distribution of the trial when the maze is fully explored (green: -ACh, brown:+ACh). C. \textbf{T-maze:} i-ii) Example trajectories. The maze has two arms, the reward is located in the right arm (with the star). i) Without cholinergic interaction (-ACh), the agent consistently goes to the the same unrewarded arm. ii) With cholinergic interaction (+ACh), the agent finds the rewarded arm sooner. iii) Percent cumulative distribution of the first rewarded trial (histogram) and the corresponding theoretical distribution (geometric distribution with $p=\frac{1}{2}$; black dots) for simulations without cholinergic depression. iv)  Percent cumulative distribution of the first rewarded trial (histogram) and the corresponding theoretical distribution (discrete uniform distribution on $[1,2]$; black dots) for agents with cholinergic depression. The empirical distribution approximates the theoretical one, but does not match it exactly. v) Percentage of successful simulations over consecutive trials for -ACh and +ACh agents (solid lines). The theoretical learning curves (assuming one-shot learning) are the cumulative distribution of the first successful trial (dashed lines with dots). vi) Agents navigating the environment without any reward. The graph shows the percent cumulative distribution of the trial when the maze is fully explored (green: -ACh, brown:+ACh). \\

\newpage
\includegraphics[width=\textwidth]{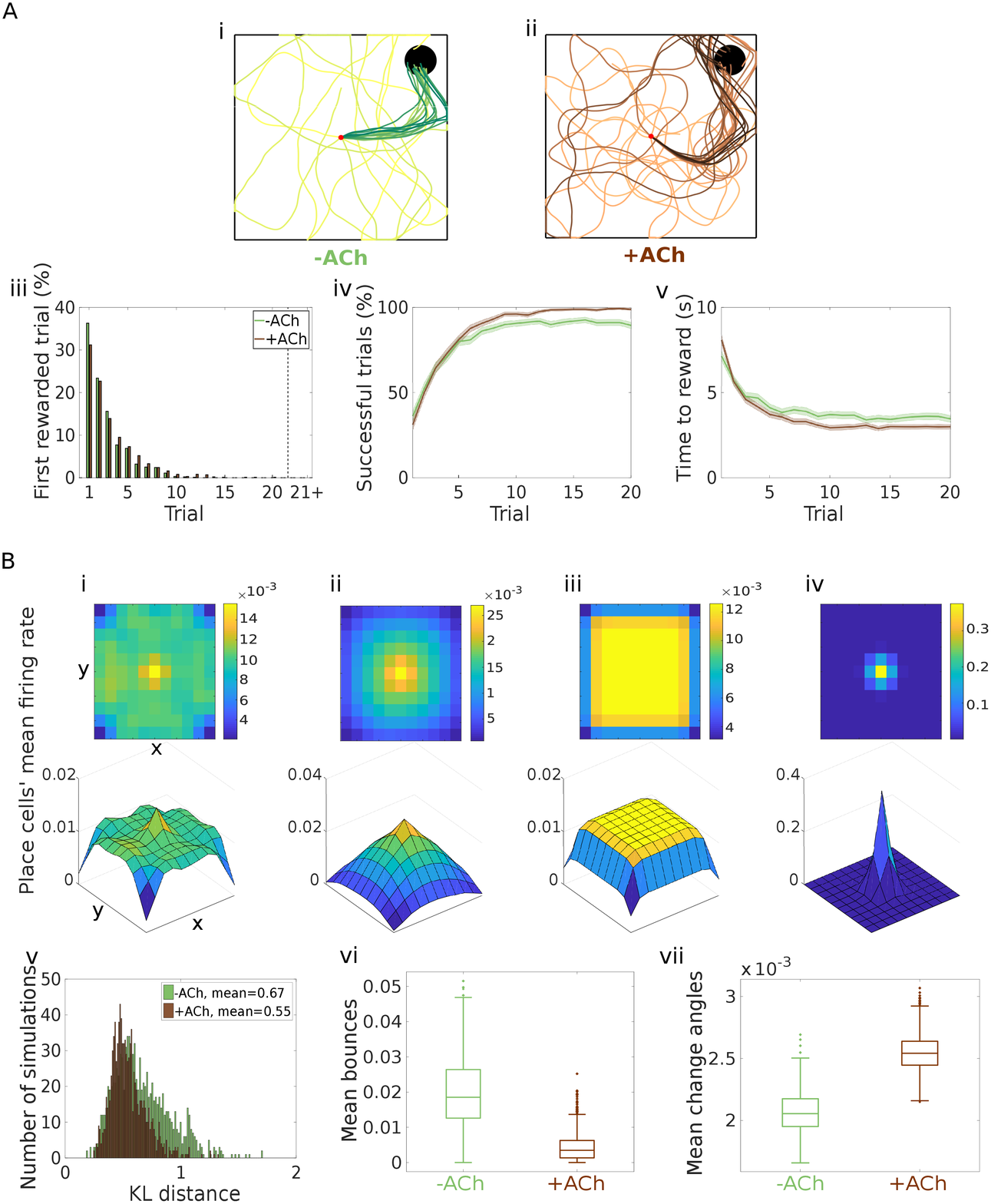}
\textbf{Fig 2. Acetylcholine-modulated plasticity enhances exploration over the action space.} A. i-ii) Example trajectories. Agents start each trial from the centre of the open field (red dot). Simulations without (-ACh, green) and with (+ACh,brown) cholinergic depression learn to navigate to the reward location (black circle) in 20 trials. Trials are coded from light to dark, according to their temporal order (early = light, late = dark).  iii) Reward discovery. Percent cumulative distribution of the first rewarded trial (-ACh, green histogram; +ACh, brown histogram).  iv) Learning curve presented as a percentage of successful simulations over successive trials. v) Average time to reward in each successful trial. Unsuccessful trials, in which the agent failed to find the reward, were excluded. B. Exploration of an open field, without any reward. i-iv) Place cells' activity during one trial, averaged across time ($T_{max} = 15$ s) and simulations ($M=1000$), displayed over the open field; 2D (top) and 3D (bottom) views. i) Simulations of the model without cholinergic depression (-ACh). ii) Simulations of the model with cholinergic depression (+ACh). iii) Benchmark simulations for exploration over the open field. Locations inside the field are sampled at random for the duration of one trial. iv) Benchmark simulations for exploration over the action space. At each timestep, actions are chosen at random, starting from the initial position. v) Histograms of the Kullback-Leibler divergences between each simulation (-ACh, green histogram; +ACh, brown histogram) and the benchmark simulations for exploration over the open field (Figure 4B.iii). vi) Boxplot of the rate of the bounces back from the walls per trial ($\frac{\text{no. of bounces}}{\text{trial duration}}$). vii) Boxplot of the difference between consecutive actions measured in radians. Acetylcholine yields greater variability in the action space. In a boxplot, the rectangle spans the 25th and 75th percentiles of the distribution. The line inside the rectangle is the median, and the whiskers indicate the minimum and the maximum.  Points outside of the box are outliers.  \\
 
\newpage
\includegraphics[width=\textwidth]{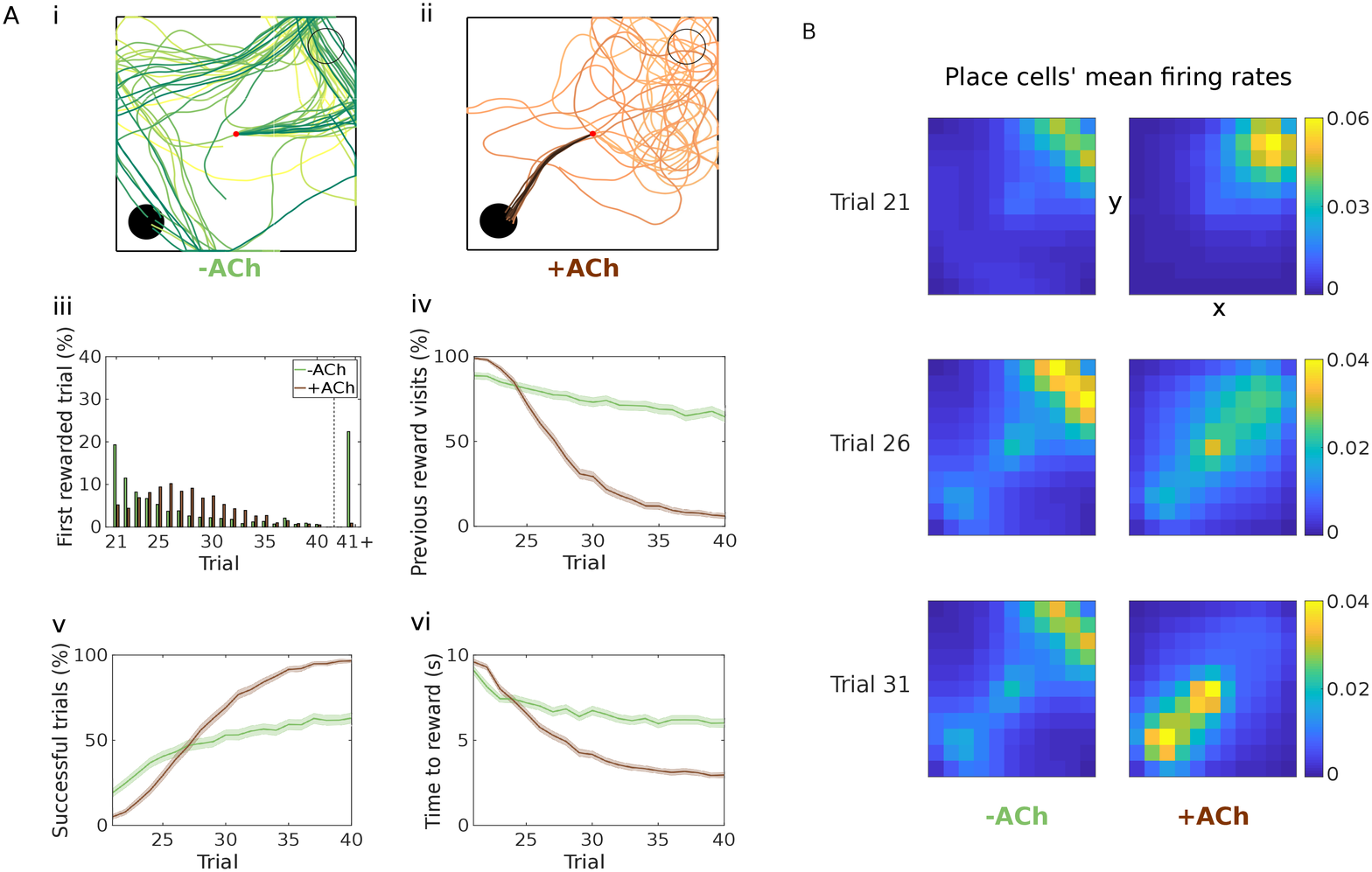} 
\textbf{Fig 3. Acetylcholine improves performance in dynamic environments.} A. After the 20 initial trials (Figure 2A), the reward is moved to the opposite corner of the open field (old location = hollow black circle, new location = solid black circle). Trials are coded from light to dark, according to their temporal order (early = light, late = dark; Trials 21-40). i) Agents without cholinergic depression do not unlearn the old path, but they can extend it to the new rewarded location. ii) Agents with cholinergic depression can forget the path learned previously. iii) Reward discovery. Percent cumulative distribution of trials when the reward is discovered for the first time. iv) Percentage of agents visiting the old reward location as a function of the trial index. v) Percentage of successful simulations as a function of the trial number. vi) Average time to reach the new reward (only successful trials). B. Place cells' activity at different trials (21, 26, 31), averaged across time and simulations, displayed as an image over the open field.\\

\newpage
\includegraphics[height=\textheight]{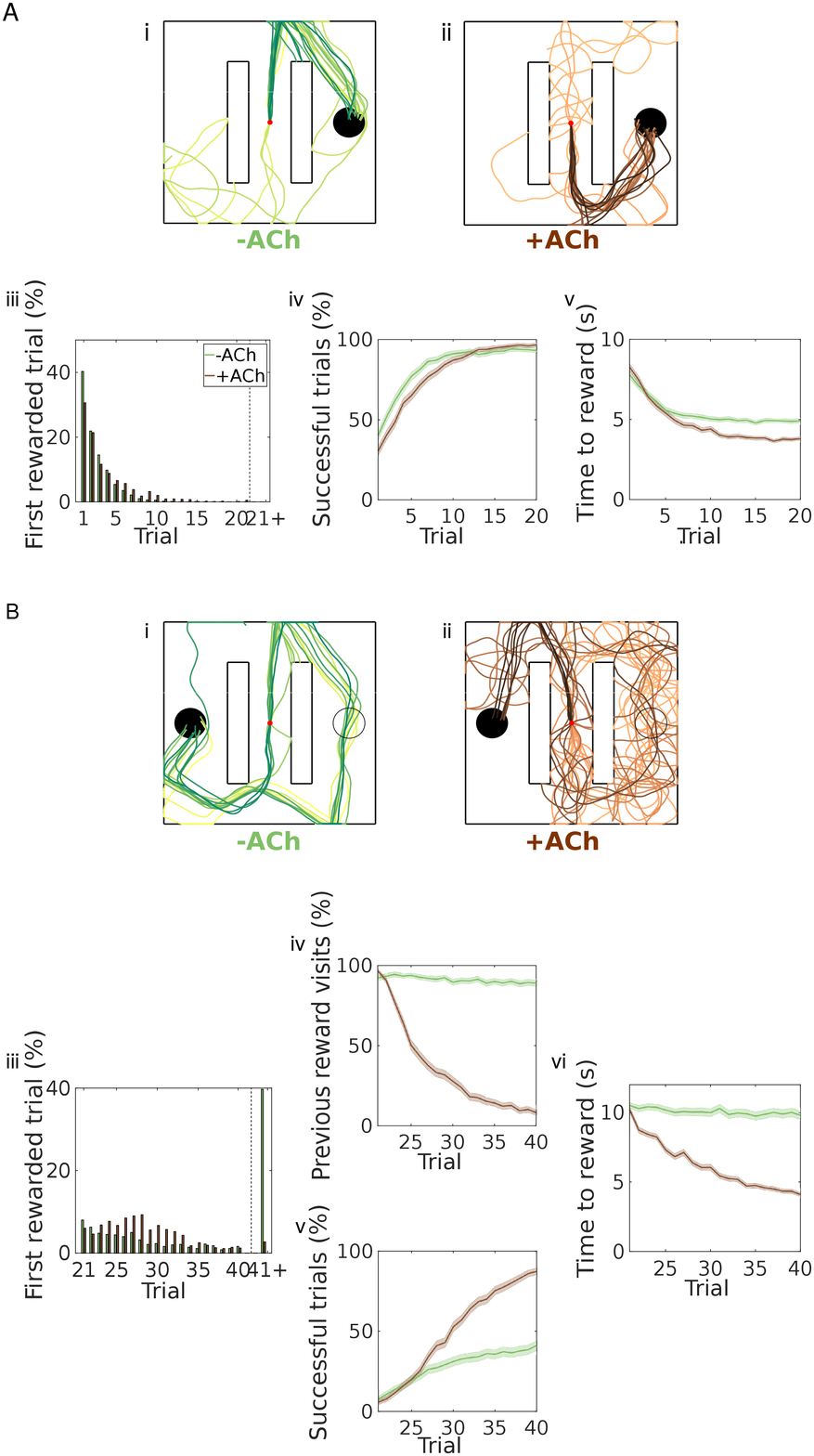} 
\textbf{Fig 4. Flexible learning in an open field with obstacles.} A. Trials 1-20.  Agents start each trial from the centre of the open field (red dot) and have to avoid the obstacles (white bars). Agents with and without cholinergic depression (+ACh and -ACh respectively) learn to navigate to the reward location (black circle). i-ii) Example trajectories. Trials are coded from light to dark, according to their temporal order (early = light, late = dark). iii) Reward discovery. Percent cumulative distribution of the first rewarded trial. iv) Percentage of successful simulations across trials. v) Average time to reward in each trial (only successful simulations). B. Trials 21-40. The reward is moved to the opposite side of the open field. Agents with and without cholinergic depression (+ACh and -ACh respectively) learn to navigate to the new reward location (solid black circle).  i) Example trajectories. Without cholinergic depression, the agent learns a route to the new reward location, but mainly as an extension of the path learned previously. ii) The agent with cholinergic depression unlearns the path to the previous reward location (hollow black circle) and navigates to the new reward. iii) Reward discovery. Percent cumulative distribution of trials when the reward is discovered for the first time. iv) Percentage of agents visiting the old reward. v) Percentage of successful simulations as a function of the trial number. vi) Average time to reach the new reward (only successful trials). \\

\newpage
\includegraphics[width=\textwidth]{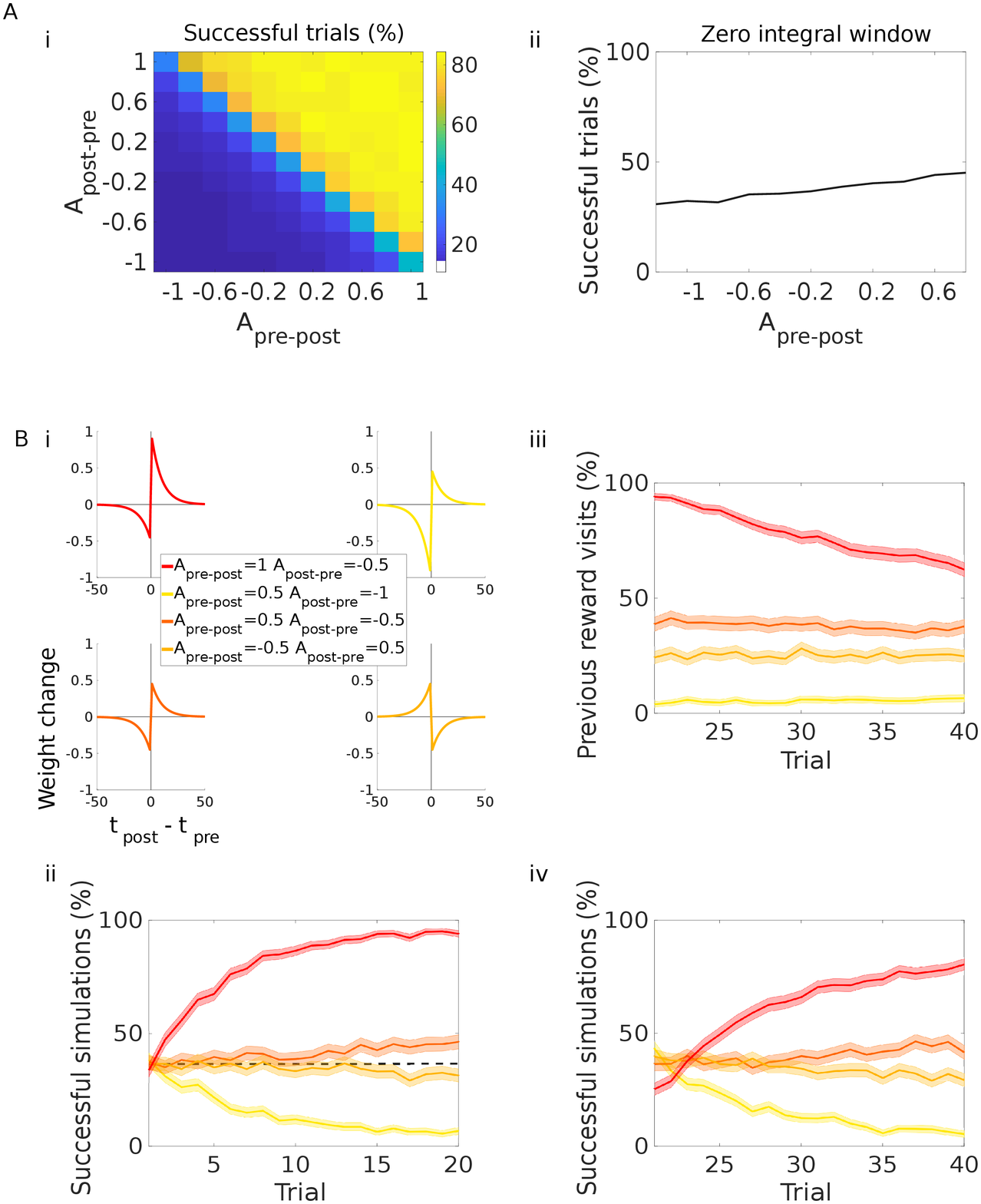} 
\textbf{Fig 5. Comparison with reward-modulated STDP.} Agents equipped with reward-modulated STDP with varying parameters $A_{pre-post}$ and $A_{post-pre}$ learn to navigate towards a reward in an open field. A: Parameters sweep. We run $M=200$ simulations of the task in Figure 2A (learning to navigate to the reward in the top-right corner of the open field, 20 trials).  i) Percentage of successful trials, as a function of the amplitudes of the plasticity windows (pre-post and post-pre). The integral of the STDP window mostly determines the agent's performance. ii) Percentage of total successful trials for an STDP window with vanishing integral ($A_{pre-post}+A_{post-pre}=0$, diagonal of the matrix). B: Simulations of the dynamic task, as in Figure 2A and 3, for four different parameters sets. i) Legend and representation of the four learning windows: positive integral (red), negative integral (yellow), zero integral with $A_{pre-post}>0$ (dark orange) and zero integral with $A_{pre-post}<0$ (light orange). ii) Agents have to learn to navigate to the reward in the top-right corner of the open field, trials 1-20. Percentage of successful simulations. The dashed line indicates the baseline performance. iii-iv) The reward is moved to the opposite corner of the open field, trials 21-40. iii) Percentage of agents visiting the previously rewarded location.  iv) Percentage of successful simulations.\\

\newpage
\includegraphics[width=\textwidth]{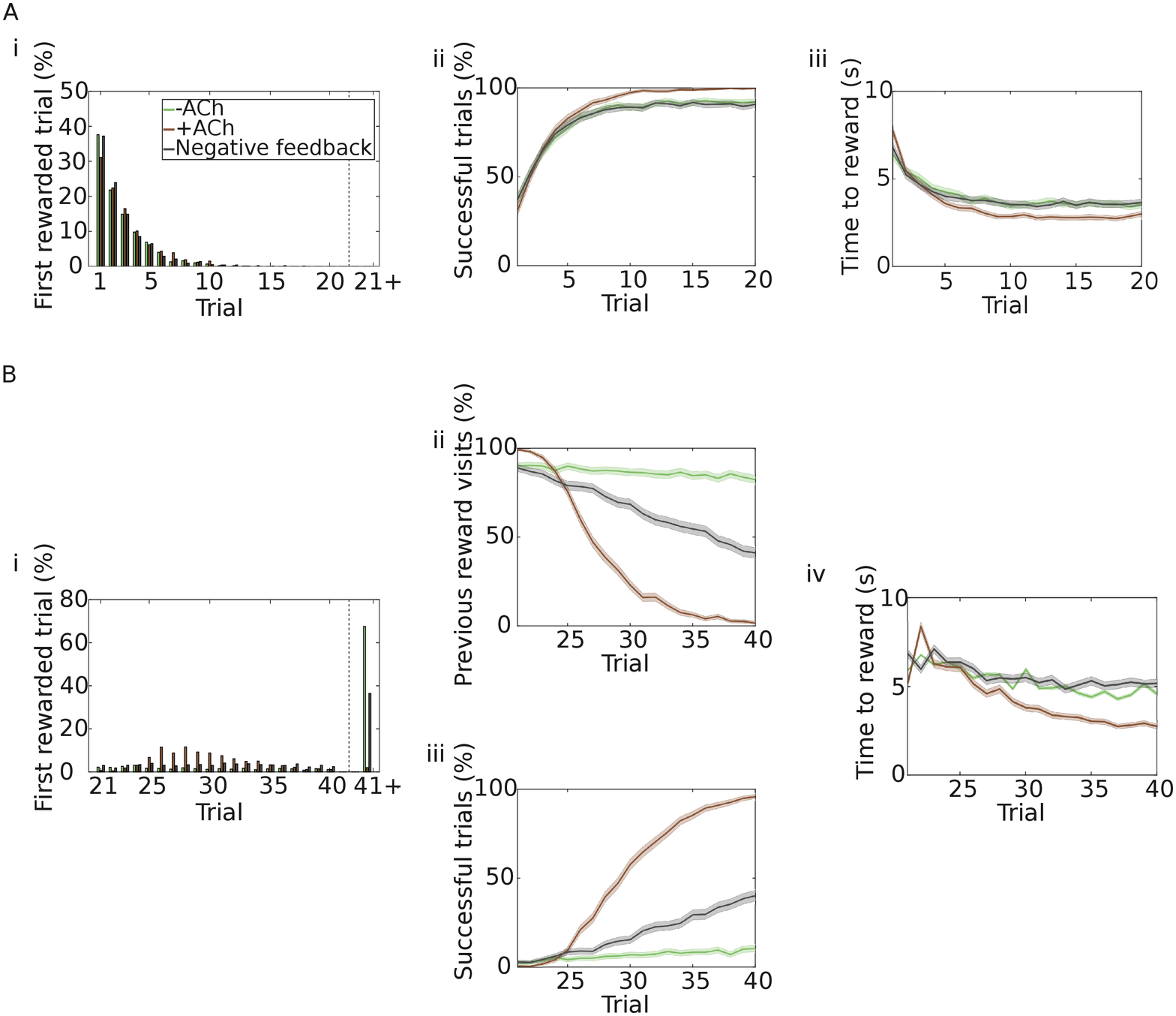} 
\textbf{Fig 6. Comparison with learning from a negative feedback.} A. Trials 1-20.  Agents start each trial from the centre of the open field and have to navigate to the top-right corner (task as in Figure 2A). Simulations are run under three different conditions: with only dopaminergic potentiation (-ACh, green); with dopaminergic potentiation and cholinergic depression (+ACh, brown); with dopaminergic potentiation and negative feedback (grey). i) Reward discovery. Percent cumulative distribution of trials when the reward is discovered for the first time. ii) Percentage of successful simulations across trials. iii) Average time to reward in each trial (only successful simulations). B. Trials 21-40. The reward is moved to the opposite side of the open field (task as in Figure 3A; but the task is stopped whenever the agent enters either the new or the old rewarded area). i) Reward discovery. Percent cumulative distribution of trials when the reward is discovered for the first time. ii) Percentage of agents visiting the old reward location. iii) Percentage of successful simulations as a function of the trial number. Acetylcholine yields the best performance.  v) Average time to reach the new reward (only successful trials). \\

\newpage
\bibliography{BibCB3}{}
\end{document}